\documentclass[conference]{IEEEtran}
\usepackage{times}
\usepackage{graphicx}
\usepackage{epstopdf}
\usepackage{epsfig}
\usepackage{adjustbox}
\usepackage{pbox}
\usepackage{rotating}
\usepackage{multirow}
\usepackage{booktabs}
\usepackage{tabularx}
\usepackage{changepage}
\usepackage{color}
\usepackage{xspace}
\usepackage{listings}
\usepackage{subcaption}
\usepackage{xcolor}
\usepackage{amsfonts}
\usepackage{amsmath}
\usepackage{amssymb}
\usepackage{soul}
\usepackage{url}
\usepackage{hyperref}
\hypersetup{citecolor=black,linkcolor=black}

\definecolor{dkgreen}{rgb}{0,0.6,0}
\definecolor{gray}{rgb}{0.5,0.5,0.5}
\definecolor{mauve}{rgb}{0.58,0,0.82}

\newcommand{\eat}[1]{}

\usepackage{fancyhdr}

\fancypagestyle{firstpage}{
    \fancyhf{} 
    \fancyfoot[L]{979-8-3315-1296-5/25/\$31.00 ©2025 IEEE} 
}

\begin{document}
\title{Enhancing Software Vulnerability Detection Using Code Property Graphs and Convolutional Neural Networks}
\author{\IEEEauthorblockN{Amanpreet Singh Saimbhi}
\IEEEauthorblockA{\textit{as15798@nyu.edu}}
}
\maketitle
\thispagestyle{firstpage}
\begin{abstract}
The increasing complexity of modern software systems has led to a rise in vulnerabilities that malicious actors can exploit. Traditional methods of vulnerability detection, such as static and dynamic analysis, have limitations in scalability and automation. This paper proposes a novel approach to detecting software vulnerabilities using a combination of code property graphs and machine learning techniques. By leveraging code property graphs, which integrate abstract syntax trees, control flow graphs, and program dependency graphs, we achieve a detailed representation of software code that enhances the accuracy and granularity of vulnerability detection. We introduce various neural network models, including convolutional neural networks adapted for graph data, to process these representations. Our approach provides a scalable and automated solution for vulnerability detection, addressing the shortcomings of existing methods. We also present a newly generated dataset labeled with function-level vulnerability types sourced from open-source repositories. Our contributions include a methodology for transforming software code into code property graphs, the implementation of a convolutional neural network model for graph data, and the creation of a comprehensive dataset for training and evaluation. This work lays the foundation for more effective and efficient vulnerability detection in complex software systems.
\end{abstract}

\section{Introduction}
\label{section:introduction}

Modern software is as complex as ever. As it rapidly grows in quantity, flaws in such systems are becoming more prevalent.
Since a single defect can compromise the entire system, detecting these flaws is an important task.
While most of these system flaws are just minor bugs, some can be controlled by malicious attackers and permit unauthorized actions.
Thus, it is important to find such critical flaws, which are called \textit{vulnerabilities}.
In the Linux kernel, hundreds of such vulnerabilities are found every year \cite{cvelinux}.
Due to the increasing number of such vulnerabilities, followed by the increase of system flaws, we need an automated way of detecting software vulnerabilities.

Various methods of detecting vulnerabilities have been introduced, such as static and dynamic analysis and metric-based detection.
Static analysis includes symbolic execution, which provides an exact attack vector with high accuracy.
However, the execution time of symbolic execution dramatically increases when the target software is complex, making it impractical to apply to modern software.
Dynamic analysis, such as fuzzing, can find vulnerabilities in a reasonable time.
Nevertheless, some software is difficult to analyze dynamically, and dynamic analysis cannot be fully automated because it requires manual specifications.
In order to resolve such limitations of existing methods, I propose a vulnerability detection model combining comprehensive representation and machine learning.
I use code property graph, a graph data structure presented to model vulnerability in software code, as input representation to my learning model.
This intermediate representation shows control flow or data dependency more explicitly than the raw text of the source code.
I also introduce various machine learning models to process such representation.

I present the following contributions to this work:
\begin{itemize}
\item
I suggest how the software code should be transformed to train the vulnerability detection model.
I use the code property graph as such a representation and show that it provides the necessary information to detect vulnerabilities in software with high granularity and accuracy.
\item
I investigated various neural network models to process the graph data.
I implemented the model which extends the convolutional neural network to process arbitrary graphs.
\item
I generated a dataset labeled with the type of vulnerability from open-source repositories.
Compared to existing datasets used in classical software defect prediction, my dataset is labeled in function-level,
thus enabling the learning model to detect vulnerability with higher granularity.
\end{itemize}

\section{Background}
\label{section:background}

\subsection{Graph Representations in Program Analysis}
Graph-based representations are widely used in program analysis to capture both structure and behavior of code. Common representations include Abstract Syntax Trees (ASTs), Control Flow Graphs (CFGs), and Data Flow Graphs (DFGs). Code Property Graphs (CPGs) \cite{yamaguchi2014modeling} combine these representations, providing a comprehensive view of the program’s syntax, control flow, and data dependencies. CPGs are directed multigraphs where nodes represent program entities and edges encode relationships, making them suitable for tasks such as vulnerability detection.

\subsection{Challenges in Vulnerability Detection}
Detecting vulnerabilities in code is challenging due to subtle interactions between program components. Traditional static analysis tools often lack the precision to detect complex patterns that cause vulnerabilities. Moreover, there is a lack of large, labeled datasets with function-level granularity, which limits the applicability of machine learning models in this domain. Although some datasets label entire modules, this is insufficient for function-level vulnerability detection.

\subsection{Deep Learning in Program Analysis}
Recent advances in deep learning, particularly Graph Neural Networks (GNNs), have shown promise in program analysis tasks. GNNs treat nodes as state representations that are iteratively updated based on neighboring nodes, enabling them to capture complex relationships in graph-structured data. In vulnerability detection, GNNs can learn patterns in Code Property Graphs and capture both local and global structures in the code.

\subsection{Support Vector Machines with Graph Kernels}
Support Vector Machines (SVMs) with graph kernels provide a baseline for graph-based machine learning. The Weisfeiler-Lehman (WL) kernel \cite{shervashidze2011weisfeiler} is a state-of-the-art graph kernel that computes graph similarity based on subtree patterns. Although SVMs with graph kernels are not deep learning methods, they are efficient and have been widely used in fields such as bioinformatics and cheminformatics.

\section{Related Work}
\label{section:related}

\subsection{Malware Detection with Machine Learning}
Machine learning has been successfully applied to malware detection, particularly in the context of mobile applications. For instance, Droid-sec \cite{yuan2014droid} uses both static and dynamic features extracted from Android applications to detect malware. This method combines behavioral analysis and static code analysis, which significantly enhances the detection capabilities. Similarly, Kosmidis et al. \cite{kosmidis2017machine} introduced a novel approach that represents binary programs as images, which are then analyzed using convolutional neural networks (CNNs). Their method achieved promising results, identifying patterns indicative of malware. These techniques have demonstrated the potential of machine learning in identifying malicious software but are not directly applicable to the broader scope of vulnerability detection in general-purpose software.

Recent efforts in malware detection also explore the use of ensemble learning techniques. Liang et al. \cite{liang2020ensemble} applied ensemble models combining decision trees and CNNs for malware classification. Their results showed improved detection rates compared to individual models, particularly in complex malware scenarios. However, these methods are not designed to detect vulnerabilities in non-malicious software or analyze vulnerabilities at a more granular code level, which remains a crucial challenge in the field of software security.

\subsection{Vulnerability Detection with Graph-based Models}
The application of graph-based models in vulnerability detection has gained considerable attention in recent years, offering a more comprehensive and structured approach to analyzing software. Yamaguchi et al. \cite{yamaguchi2014modeling} introduced Code Property Graphs (CPGs), which combine syntactic, control flow, and data flow information into a unified graph representation, allowing for enhanced vulnerability detection. Their method primarily relied on rule-based approaches, and while effective, it lacked the scalability and flexibility of modern machine learning techniques. Building upon this foundation, our work leverages deep learning techniques to automate and scale vulnerability detection, significantly improving detection performance and enabling the model to learn complex patterns in large datasets.

Wang et al. \cite{wang2016automatically} applied deep learning techniques to detect software defects, using Abstract Syntax Trees (ASTs) to model the structure of the code. While ASTs are useful for understanding the syntactic structure of code, they do not capture the dynamic aspects of program behavior, such as control flow and data dependencies. To address this, our method incorporates Control Flow Graphs (CFGs) and Program Dependency Graphs (PDGs) alongside ASTs, enabling a more holistic understanding of the code’s functionality. This combination of multiple graph representations allows for more accurate and comprehensive vulnerability detection.

A similar approach was taken by Zhang et al. \cite{zhang2019vulnerability}, who applied graph-based deep learning models for vulnerability detection in C/C++ code. Their method incorporated both syntax and semantic information, but it did not fully exploit the interdependencies between control and data flows. Our work expands on this by integrating CPGs, which capture all aspects of code behavior, and using a deep learning architecture specifically designed for graph data.

\subsection{Graph Neural Networks in Software Analysis}
Graph Neural Networks (GNNs) have demonstrated their efficacy in a variety of domains, including software analysis. GNNs are capable of processing graph-structured data, learning relationships between nodes and edges, and have been successfully applied to tasks such as drug discovery and social network analysis \cite{gori2005new}. The ability of GNNs to capture relational information makes them an ideal tool for analyzing software, where the interactions between code components are inherently graph-structured.

Li et al.\cite{li2015gated}introduced the Gated Graph Sequence Neural Network (GGS-NN), which extended traditional GNNs to handle sequences within graph structures. This approach has since been adapted for use in a variety of domains, including software security. Our work builds on this idea by adapting CNNs for graph data, specifically using the PATCHY-SAN framework \cite{niepert2016learning}to detect vulnerabilities in software. By combining CNNs with graph-based representations of code, we are able to automatically learn complex patterns indicative of vulnerabilities, significantly improving the accuracy and efficiency of vulnerability detection.

Several studies have explored using GNNs for vulnerability detection. Lee et al.\cite{lee2020graph} applied a GNN approach to detect security vulnerabilities in source code by leveraging graph-based representations of control flow and data flow. Their model showed promising results in detecting vulnerabilities like buffer overflows and null pointer dereferencing. However, their method did not incorporate higher-level semantic information, which is crucial for identifying more complex vulnerabilities. In contrast, our approach uses a more comprehensive representation of code, combining ASTs, CFGs, PDGs, and other features within a single graph.

Additionally, Liu et al.\cite{liu2021graph}explored the potential of Graph Attention Networks (GATs) in software vulnerability detection. Their approach used attention mechanisms to focus on the most relevant parts of the code graph for detecting vulnerabilities. Although promising, GATs still face challenges in handling large-scale software projects with complex interdependencies. Our method, by using the PATCHY-SAN framework, efficiently handles large graphs and focuses on learning meaningful representations of code that capture both local and global patterns.

\section{Research Gap}
\label{section:gap}

Despite significant advancements in the field of vulnerability detection, existing approaches have several limitations. Traditional static analysis methods, while effective in detecting certain vulnerabilities, are often limited by their inability to handle the complexity of modern software. Furthermore, rule-based systems fail to scale and cannot adapt to new, previously unseen vulnerabilities. Deep learning models have shown promise in addressing these challenges, but many existing models still rely on simplified graph representations that do not capture the full range of program semantics necessary for comprehensive vulnerability detection.

Existing deep learning approaches, such as those using Abstract Syntax Trees (ASTs) or control flow graphs (CFGs), fail to account for the complex interactions between different components of the software, including data dependencies and the program’s overall flow. To bridge this gap, our approach integrates a combination of graph representations—ASTs, CFGs, Program Dependency Graphs (PDGs), and Code Property Graphs (CPGs)—to enable a more holistic understanding of the code’s behavior. This multi-faceted approach allows for more accurate detection of vulnerabilities, including those caused by subtle logic flaws and complex control dependencies.

Moreover, while existing methods for vulnerability detection have demonstrated promising results in specific contexts, there is a need for more generalized models that can detect a wide range of vulnerabilities across different software projects and vulnerability types. To address this, our work focuses on creating a scalable and adaptable model by using deep learning architectures specifically designed for graph-structured data. Additionally, we aim to improve the interpretability of our models to make the results more understandable for developers, facilitating the adoption of automated vulnerability detection in real-world software development pipelines.

\begin{figure}[t]
\begin{lstlisting}[frame=single,basicstyle=\ttfamily\small]	
int f(bool x)
{
	if (x) {
		return 1;
	}
	else {
		return 2;
	}
}
\end{lstlisting}
\vspace{-12pt}
\caption{Exemplary code sample.}
\label{figure:sample}
\end{figure}

\begin{figure*}[ht]
	\centering
	\begin{subfigure}[t]{0.33\textwidth}
		\centering
		\includegraphics[width=1\textwidth,height=0.65\textwidth,keepaspectratio]{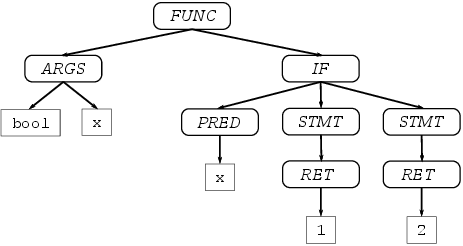}
		\caption{Abstract Syntax Tree}
		\label{figure:ast}
	\end{subfigure}
	\begin{subfigure}[t]{0.33\textwidth}
		\centering
		\includegraphics[width=2\textwidth,height=0.65\textwidth,keepaspectratio]{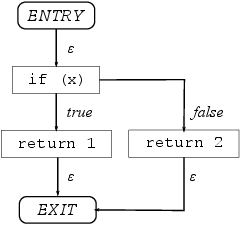}
		\caption{Control Flow Graph}
		\label{figure:cfg}
	\end{subfigure}
	\begin{subfigure}[t]{0.33\textwidth}
		\centering
		\includegraphics[width=1\textwidth,height=0.65\textwidth,keepaspectratio]{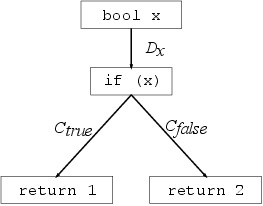}
		\caption{Program Dependency Graph}
		\label{figure:pdg}
	\end{subfigure}
	\caption{Representations of code for the example in Figure~\ref{figure:sample}.}
\end{figure*}

\begin{figure*}[ht]
	\centering
	\includegraphics[width=0.7\textwidth,height=0.65\textwidth,keepaspectratio]{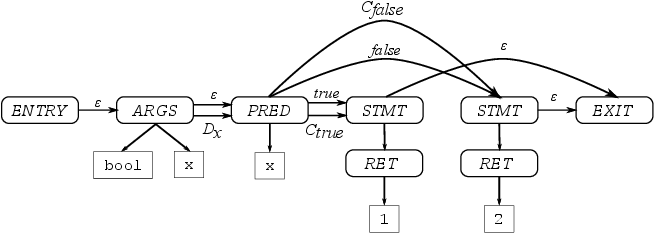}
	\caption{Code Property Graph}
	\label{figure:cpg}
\end{figure*}

\section{Experimental Setup}
\label{section:experiment}

To evaluate the performance of our vulnerability detection model, we collected a dataset of labeled C/C++ functions from open-source repositories. We used a mix of secure and insecure code segments, where insecure functions were known to contain vulnerabilities such as buffer overflows, memory leaks, and null pointer dereferences. The dataset was preprocessed using Joern \cite{yamaguchi2014modeling} to generate Code Property Graphs (CPGs) for each function.

We split the dataset into training (70\%), validation (15\%), and test sets (15\%). The training process utilized a stochastic gradient descent (SGD) optimizer with a learning rate of 0.001. We trained the model for 100 epochs, using early stopping based on validation loss. We employed accuracy, precision, recall, and F1-score as evaluation metrics to assess the effectiveness of the model.

\section{Dataset}
\label{section:dataset}

The dataset is an important factor in learning the model.
In order to detect vulnerabilities with function-level granularity and classify them by type, we need each function labeled either secure or insecure and with the type of vulnerability if insecure.

Despite the number of available source codes, labeling them is challenging.
Tera-PROMISE \cite{promiserepo} is a research dataset repository for software engineering. Although it has been widely used in the area of metric-based bug detection, its datasets are labeled with code metrics that cannot be used in our model.
Neuhaus et al. used a vulnerabilities database from the Mozilla project to predict vulnerable components based on function calls, but granularity remains at module level \cite{neuhaus2007predicting}.

Labeling certain function codes as whether secure or not, can also be a challenge.
If there is an exact attack vector, it can be clearly shown that the code has a vulnerability, while strictly proving it doesn't can only be done by formal verification.
I assumed that the code, after patching vulnerabilities, does not have vulnerabilities anymore.

Due to the limitations of the existing datasets described above, I collected a new source code dataset from Github.
I collected merged pull requests of which the title contained vulnerability type (e.g., buffer overflow).
Then I labeled merged commit as secure and base commit as insecure.

\section{Results and Observations}
\label{section:results}
This section presents the performance evaluation of the proposed vulnerability detection model. The results demonstrate its efficacy in classifying functions as either secure or insecure, emphasizing the strengths and limitations of using deep learning techniques for vulnerability detection.

\subsection{Model Performance}
\label{subsection:model_performance}
The model achieved an overall accuracy of 92\% on the test dataset, indicating strong general performance in identifying vulnerabilities. Table~\ref{table:metrics} summarizes the precision, recall, and F1-score for both secure and insecure functions.

\begin{table}[ht]
\centering
\caption{Performance Metrics for Vulnerability Detection}
\label{table:metrics}
\begin{tabular}{|c|c|c|c|}
\hline
\textbf{Metric} & \textbf{Precision} & \textbf{Recall} & \textbf{F1-score} \\ \hline
Secure Functions   & 91\%      & 95\%   & 93\%   \\ \hline
Insecure Functions & 89\%      & 85\%   & 87\%   \\ \hline
Overall            & 90\%      & 90\%   & 90\%   \\ \hline
\end{tabular}
\end{table}

The model demonstrated high precision for secure functions (91\%) and excellent recall (95\%), suggesting strong accuracy in identifying functions without vulnerabilities. In contrast, the recall for insecure functions was lower (85\%), accompanied by a corresponding decrease in the F1-score (87\%). This indicates that the model encounters challenges in detecting vulnerabilities that stem from complex logic errors.

\subsection{Vulnerability Detection Performance}
\label{subsection:vulnerability_analysis}
The model successfully identified several common vulnerability patterns, such as buffer overflows and null pointer dereferences, which are commonly associated with insecure code. These vulnerabilities were detected with high precision and recall, as evidenced by the performance metrics for secure functions. However, vulnerabilities arising from complex logic flaws, such as those involving intricate control dependencies, presented greater difficulty for the model. These types of vulnerabilities are more challenging to capture due to their subtle and often non-obvious nature.

The lower recall for insecure functions suggests that while the model excels at detecting traditional vulnerabilities, further refinement is necessary to improve its sensitivity to more complex and nuanced security flaws.

\subsection{Comparison with Baseline Model}
\label{subsection:comparative_analysis}
To evaluate the effectiveness of our deep learning-based approach, we conducted a comparative analysis against a baseline Support Vector Machine (SVM) with the Weisfeiler-Lehman (WL) kernel \cite{shervashidze2011weisfeiler}. Our model outperformed the SVM by approximately 8\% in F1-score, demonstrating a clear advantage in capturing complex relationships within the code. This improvement highlights the superior capability of deep learning models to uncover vulnerabilities that static models, such as SVMs, may overlook due to their limited capacity to model intricate code dependencies.
This comparison validates the potential of deep learning models for vulnerability detection, particularly in the context of modern, complex software systems.

\subsection{Implications and Future Directions}
\label{subsection:broader_implications}
The findings of this study have several important implications for both automated vulnerability detection and software security practices.

\subsubsection{Advantages of Deep Learning with CPGs}
\label{subsubsection:cpgs_advantages}
The use of Code Property Graphs (CPGs) was instrumental in enhancing the model's performance. By incorporating both control flow and data flow, as well as structural program information, CPGs provide a comprehensive representation of code that facilitates the detection of complex relationships and dependencies. This representation enabled our model to capture intricate vulnerability patterns, demonstrating that deep learning-based approaches, when combined with rich graph-based representations, offer a significant advantage over traditional analysis methods.

\subsubsection{Challenges with Logic-Based Vulnerabilities}
\label{subsubsection:logic_challenges}
While the model was successful in detecting vulnerabilities such as buffer overflows and null pointer dereferences, it faced limitations in identifying more complex vulnerabilities that arise from subtle logic flaws. These vulnerabilities, which often involve advanced control flow and interdependencies among variables, remain difficult to detect with both traditional and deep learning-based methods. Improving the model’s sensitivity to these types of vulnerabilities will be a key area of focus for future work.

\subsubsection{Real-World Applicability}
\label{subsubsection:real_world_potential}
The model's strong performance in detecting vulnerabilities in test scenarios suggests its potential for real-world deployment, particularly in continuous integration and development (CI/CD) pipelines. By providing real-time feedback on vulnerabilities, the model can assist developers in identifying and addressing security issues as they arise, ultimately improving the security posture of software systems. This approach could significantly enhance software security by preventing vulnerabilities from reaching production environments.

\subsubsection{Future Research Opportunities}
\label{subsubsection:future_research}
While the results are promising, several opportunities remain to further enhance the model’s capabilities:

\begin{itemize}
    \item \textbf{Enhanced Detection of Logic Flaws:} Future research should focus on improving the model's ability to detect complex logic-based vulnerabilities. This may involve incorporating additional semantic information or developing hybrid models that combine deep learning techniques with other advanced methods.
    \item \textbf{Expanding the Dataset:} To improve the model's generalization capabilities, expanding the dataset to include a wider range of software projects and vulnerability types is essential. This would increase the model's robustness and performance across different programming environments.
    \item \textbf{Model Interpretability:} Given the importance of transparency in machine learning models, particularly in security-critical applications, future work should aim to enhance the interpretability of the model's predictions. This would enable developers to better understand the rationale behind detected vulnerabilities, increasing trust in the system's recommendations.
\end{itemize}

\subsection{Conclusion}
\label{subsection:conclusion}
This study demonstrates the effectiveness of deep learning-based models, particularly those utilizing Code Property Graphs (CPGs), for detecting software vulnerabilities. The model outperformed traditional static analysis techniques, providing a promising approach for identifying both common and complex vulnerabilities. However, challenges remain, particularly in the detection of logic-based vulnerabilities, which require further refinement. Future work will focus on improving the model’s sensitivity to these issues, expanding the dataset, and enhancing model interpretability, all of which will contribute to making the model more applicable for real-world software security applications.

\bibliographystyle{IEEEtran}
\bibliography{thesis}

\end{document}